# Hierarchical Modeling of Multifunctional Novel Carbon Nanotube Reinforced Hybrid Composites for Next Generation Polymeric Composites

S. I. KUNDALWAL

*Department of Mechanical Engineering, Indian Institute of Technology, Indore, 453552, India*

Email: Kundalwal@iiti.ac.in

## ABSTRACT

*This article provides an overview of the modeling of the effective thermomechanical properties of the multifunctional carbon nanotube (CNT) reinforced hybrid composites for advanced structural applications. The novel constructional feature of such multifunctional CNT-reinforced hybrid composite is that CNTs are radially grown on the circumferential surfaces of the carbon fiber reinforcements. Several micromechanical models have been developed to predict the effective thermomechanical properties of these multifunctional CNT-reinforced hybrid composites. The waviness of CNTs is intrinsic to many manufacturing processes and influences the thermomechanical behavior of CNT-reinforced composites. An endeavor has also been made to investigate the effect of wavy CNTs on the thermomechanical properties of the multifunctional CNT-reinforced hybrid composites. Radially aligned CNTs grown on the circumferential surfaces of the carbon fibers assure that the CNT-reinforced hybrid composites are truly multifunctional and may be a promising advanced next generation composite for structural applications.*

# 1. INTRODUCTION

The discovery of carbon nanotubes (CNTs) [1] has stimulated a tremendous research on the prediction of their remarkable mechanical and thermal properties. Researchers probably thought that CNTs may be useful as nanoscale fibers for developing novel CNT-reinforced nanocomposites and this conjecture motivated them to accurately predict their thermomechanical properties. Numerous experimental and numerical studies revealed that the axial Young's modulus of CNTs is in the terapascal range [2–5]. As nanoscale graphite structures, CNTs are of great interest not only for their mechanical properties but also for their thermal properties. CNTs exhibit thermal properties that are remarkably different from other known materials and are expected to be a promising candidate in many advanced applications [6–8]. The quest for utilizing such exceptional thermomechanical properties of CNTs and their high aspect ratio led to the opening of an emerging area of research on the development of CNT-reinforced nanocomposites [5, 9–12]. However, manufacturing of such two-phase nanocomposites reinforced with long CNTs has some challenging technical issues such as waviness, agglomeration, misalignment, adhesion of CNTs in polymer matrix and difficulty in manufacturing long CNTs [13, 14]. These difficulties can be alleviated by using CNTs as secondary reinforcements in the three-phase CNT-reinforced hybrid composites. In case of three-phase CNT-reinforced hybrid composite, CNTs are grown on the circumferential surfaces of the advanced fiber reinforcements [15, 16]. Such a fiber augmented with radially grown CNTs on its circumferential surface is being called as "fuzzy fiber" [15] and the resulting composite is called as fuzzy fiber reinforced composite (FFRC) [16-27].

As reported in the open literature, researchers mainly devoted to investigate the enhancement of tensile strength, interfacial shear strength, fracture toughness, impact resistance of such CNT-reinforced hybrid composites. Prediction of all effective thermomechanical properties of these CNT-reinforced hybrid composites appears to be an important issue for further research. In order to establish these CNT-reinforced hybrid composites as the superior advanced composite for structural applications, structural analysis must be carried out using these CNT-reinforced hybrid composites and for such analysis all effective properties of this composite must be known a priori. However, studies concerning the estimation of all effective properties of the CNT-reinforced hybrid composites have not yet been reported in the open literature. Such lack in studies provides an ample scope for further research on developing accurate models for predicting all effective properties of the multifunctional CNT-reinforced hybrid composites. Hence, the present study is directed to estimate all effective thermomechanical properties of such multifunctional CNT-reinforced hybrid composites.

It has been experimentally observed that CNTs are actually curved cylindrical tubes with a relatively high aspect ratio [28–30]. Therefore, the effect of waviness of CNTs on the effective properties of the CNT-reinforced hybrid composites is also investigated in the present study when the wavy CNTs are coplanar with either of the two mutually orthogonal planes.

## 2. NOVEL FUZZY FIBER REINFORCED COMPOSITES

The schematic diagram illustrated in Fig. 1 represents a lamina of the FFRC being studied here. In this novel composite, the wavy CNTs are radially grown on the circumferential surfaces of the carbon fiber reinforcements while they are uniformly spaced on the circumferential surfaces of the carbon fibers. Such a resulting fuzzy fiber coated with the wavy CNTs is shown in Fig. 2. In the present study, the wavy CNTs are modeled as sinusoidal solid CNT fibers [17–20] while at any location along the length of the CNT, the CNT is considered as transversely isotropic [4, 5, 11]. The polymer matrix is reinforced by the fuzzy fiber coated with the wavy CNTs and such combination can be viewed as a circular cylindrical composite fuzzy fiber (CFF) in which the carbon fiber is embedded in the wavy CNT-reinforced polymer matrix nanocomposite (PMNC). It may be noted that the variations of the constructional feature of the CFF can be such that the wavy CNTs are coplanar with the 2–3 (2'–3') plane or the 1–3 (1'–3') plane as shown in Figs. 3 (a) and 3 (b), respectively. In case of the wavy CNTs being coplanar with the 2–3 (2'–3') plane, the amplitudes of the CNT waves are transverse to the axes of carbon fibers (i.e., 1–direction) while the amplitudes of the wavy CNTs being coplanar with the 1–3 (1'–3') plane are parallel to the axes of carbon fibers.

### 2.1 Models of the wavy CNTs

Considering a carbon fiber and an unwound lamina while this lamina is composed of the sinusoidally wavy CNTs which are coplanar with either the 2–3 (2'–3') plane or the 1–3 (1'–3') plane, the CFF can be viewed to be formed by wrapping the carbon fiber with the unwound lamina of the PMNC as shown in Fig. 4. The RVE of the unwound PMNC material containing a wavy CNT has been illustrated in Fig. 5. As shown in Fig. 5, the RVE is divided into infinitesimally thin slices of thickness dy. Averaging the effective properties of these slices over the length ($L_n$) of the RVE (i.e., the thickness of the unwound lamina of the PMNC), the homogenized effective properties of the unwound PMNC can be estimated. Each slice can be treated as an off-axis unidirectional lamina and its effective properties can be determined by transforming the effective properties of the corresponding orthotropic lamina. Now, these wavy CNTs are characterized by

$$z = A\sin(\omega y) \quad \text{or} \quad x = A\sin(\omega y) \; ; \quad \omega = n\pi/L_n \tag{1}$$

according as the wavy CNTs are coplanar with the 2–3 (2'–3') plane or the 1–3 (1'–3') plane, respectively. In Eq. (1), A and $L_n$ are the amplitude of the CNT wave and the linear distance between the CNT ends, respectively, and n represents the number of waves of the CNT. The running length ($L_{nr}$) of the CNT can be expressed in the following form:

$$L_{nr} = \int_0^{L_n} \sqrt{1 + A^2\omega^2\cos^2(\omega y)} \; dy \tag{2}$$

in which the angle ϕ shown in Fig. 5 is given by

$$\tan\phi = dz/dy = A\omega \cos(\omega y) \quad \text{or} \quad \tan\phi = dx/dy = A\omega \cos(\omega y) \tag{3}$$

according as the wavy CNT is coplanar with the 2–3 (2'–3') or the 1–3 (1'–3') plane, respectively. Note that for a particular value of ω, the value of ϕ varies with the amplitude of the CNT wave.

## 2.2 Effective thermoelastic properties of the FFRC

This Section deals with the procedures of employing two modeling approaches, namely, the Mori-Tanaka (MT) method and the mechanics of materials (MOM) approach to predict the effective thermoelastic properties of the FFRC. The RVE of the FFRC can be treated as being composed of the two phases wherein the reinforcement is the CFF and the matrix is the polymer material. Thus the analytical procedure for estimating the effective thermoelastic properties of the FFRC starts with the estimation of the effective thermoelastic properties of the PMNC containing wavy CNTs. Subsequently, considering the PMNC material as the matrix phase and the carbon fiber as the reinforcement, effective thermoelastic properties of the CFF are to be computed. Finally, using the thermoelastic properties of the CFF and the polymer matrix, the effective thermoelastic of the FFRC can be estimated.

It may be noted that the effective thermoelastic properties at any point in the unwound lamina of the PMNC containing sinusoidally wavy CNTs where the CNT axis makes an angle ϕ with the 3 (3')–axis can be approximated by transforming the effective thermoelastic properties of the unwound lamina of the PMNC containing straight CNTs. Hence, in what follows the method of deriving the MT model for predicting the effective thermoelastic properties of the unwound lamina of the PMNC containing straight CNTs will be presented first. Utilizing the effective elastic properties of the CNT and the polymer matrix properties, the MT model [31] can be derived to estimate the effective elastic coefficient matrix $[C^{nc}]$ of the unwound PMNC. The explicit formulation of the MT model for the unwound PMNC material can be derived as

$$[C^{nc}] = [C^p] + v_n([C^n] - [C^p])[\tilde{A}_1]\left[v_p[I] + v_n[\tilde{A}_1]\right]^{-1} \quad (4)$$

in which the matrix of the strain concentration factors is given by

$$[\tilde{A}_1] = \left[[I] + [S_1]([C^p])^{-1}([C^n] - [C^p])\right]^{-1} \quad (5)$$

where $v_n$ and $v_p$ represent the volume fractions of the CNT fiber and the polymer material, respectively, present in the RVE of the PMNC while $[S_1]$ represents the Eshelby tensor and the specific form of the Eshelby tensor for cylindrical inclusion given by Qui and Weng [32] is utilized. Using the effective elastic coefficient matrix $[C^{nc}]$, the effective thermal expansion coefficient vector $\{\alpha^{nc}\}$ for the unwound PMNC material can be derived in the form [33] as follows:

$$\{\alpha^{nc}\} = \{\alpha^n\} + ([C^{nc}]^{-1} - [C^n]^{-1})([C^n]^{-1} - [C^p]^{-1})^{-1}(\{\alpha^n\} - \{\alpha^p\}) \quad (6)$$

where $\{\alpha^n\}$ and $\{\alpha^p\}$ are the thermal expansion coefficient vectors of the CNT fiber and the polymer material, respectively. The effective elastic coefficients ($C_{ij}^{NC}$) and the effective thermal expansion coefficients ($\alpha_{ij}^{NC}$) at any point in the unwound lamina of the PMNC where the CNT is inclined at an angle ϕ with the 3 (3')–axis can be derived in a straightforward manner by employing the appropriate transformation law. Thus if the plane of the CNT waviness is coplanar with the 2–3 (2'–3') plane, the effective elastic ($C_{ij}^{NC}$) and thermal expansion ($\alpha_{ij}^{NC}$) coefficients at any point in the unwound lamina of the PMNC are given by

$$C_{11}^{NC} = C_{11}^{nc}, \quad C_{12}^{NC} = C_{12}^{nc}k^2 + C_{13}^{nc}l^2, \quad C_{13}^{NC} = C_{12}^{nc}l^2 + C_{13}^{nc}k^2,$$

$$C_{22}^{NC} = C_{22}^{nc}k^4 + C_{33}^{nc}l^4 + 2(C_{23}^{nc} + 2C_{44}^{nc})k^2l^2,$$

$$C_{23}^{NC} = (C_{22}^{nc} + C_{33}^{nc} - 4C_{44}^{nc})k^2l^2 + C_{23}^{nc}(k^4 + l^4),$$

$$C_{33}^{NC} = C_{22}^{nc}l^4 + C_{33}^{nc}k^4 + 2(C_{23}^{nc} + 2C_{44}^{nc})k^2l^2,$$

$$C_{44}^{NC} = (C_{22}^{nc} + C_{33}^{nc} - 2C_{23}^{nc} - 2C_{44}^{nc})k^2l^2 + C_{44}^{nc}(k^4 + l^4),$$

$$C_{55}^{NC} = C_{55}^{nc}k^2 + C_{66}^{nc}l^2, \quad C_{66}^{NC} = C_{55}^{nc}l^2 + C_{66}^{nc}k^2,$$

$$\alpha_{11}^{NC} = \alpha_{11}^{nc}, \quad \alpha_{22}^{NC} = \alpha_{22}^{nc}k^2 + \alpha_{33}^{nc}l^2 \quad \text{and} \quad \alpha_{33}^{NC} = \alpha_{22}^{nc}l^2 + \alpha_{33}^{nc}k^2 \tag{7}$$

in which

$$k = \cos\phi = [1 + \{n\pi A/L_n \cos(n\pi y/L_n)\}^2]^{-1/2} \quad \text{and}$$

$$l = \sin\phi = n\pi A/L_n \cos(n\pi y/L_n)\,[1 + \{n\pi A/L_n \cos(n\pi y/L_n)\}^2]^{-1/2}$$

Similarly, if the plane of the CNT waviness is coplanar with the 1–3 (1'–3') plane, then the effective elastic ($C_{ij}^{NC}$) and thermal expansion ($\alpha_{ij}^{NC}$) coefficients at any point of the unwound PMNC lamina where the CNT is inclined at an angle ϕ with the 3 (3') –axis are given by

$$C_{11}^{NC} = C_{11}^{nc}k^4 + C_{33}^{nc}l^4 + 2(C_{13}^{nc} + 2C_{55}^{nc})k^2l^2, \quad C_{12}^{NC} = C_{12}^{nc}k^2 + C_{23}^{nc}l^2,$$

$$C_{13}^{NC} = (C_{11}^{nc} + C_{33}^{nc} - 4C_{55}^{nc})k^2l^2 + C_{13}^{nc}(k^4 + l^4), \quad C_{22}^{NC} = C_{22}^{nc},$$

$$C_{23}^{NC} = C_{12}^{nc}l^2 + C_{23}^{nc}k^2, \quad C_{33}^{NC} = C_{11}^{nc}l^4 + C_{33}^{nc}k^4 + 2(C_{13}^{nc} + 2C_{55}^{nc})k^2l^2,$$

$$C_{44}^{NC} = C_{44}^{nc}k^2 + C_{66}^{nc}l^2, \quad C_{55}^{NC} = (C_{11}^{nc} + C_{33}^{nc} - 2C_{13}^{nc} - 2C_{55}^{nc})k^2l^2 + C_{55}^{nc}(k^4 + l^4),$$

$$C_{66}^{NC} = C_{44}^{nc}l^2 + C_{66}^{nc}k^2, \ \alpha_{11}^{NC} = \alpha_{11}^{nc}k^2 + \alpha_{33}^{nc}l^2, \ \alpha_{22}^{NC} = \alpha_{22}^{nc} \ \text{and} \ \alpha_{33}^{NC} = \alpha_{11}^{nc}l^2 + \alpha_{33}^{nc}k^2 \tag{8}$$

It is now obvious that the effective thermoelastic properties of the unwound PMNC lamina with the wavy CNTs vary along the length of the CNT as the value of ϕ vary over the length of the CNT. The average effective elastic coefficient matrix $[\bar{C}^{NC}]$ and the average thermal expansion coefficient vector $\{\bar{\alpha}^{NC}\}$ of the lamina of such unwound PMNC material containing wavy CNTs can be obtained by averaging the transformed elastic $(C_{ij}^{NC})$ and thermal expansion $(\alpha_{ij}^{NC})$ coefficients over the linear distance between the CNT ends as follows [34]:

$$[\bar{C}^{NC}] = \frac{1}{L_n}\int_0^{L_n}[C^{NC}]\,dy \quad \text{and} \quad \{\bar{\alpha}^{NC}\} = \frac{1}{L_n}\int_0^{L_n}\{\alpha^{NC}\}\,dy \tag{9}$$

It may also be noted that when the carbon fiber is viewed to be wrapped by such unwound lamina of the PMNC, the matrix $[\bar{C}^{NC}]$ and the vector $\{\bar{\alpha}^{NC}\}$ provides the effective properties at a point located in the PMNC where the CNT axis (3'–axis) is oriented at an angle θ with the 3–axis in the 2–3 plane as shown in Fig. 3 and 4. Hence, at any point in the PMNC surrounding the carbon fiber, the effective elastic coefficient matrix $[\bar{C}^{PMNC}]$ and the effective thermal expansion coefficient vector $\{\bar{\alpha}^{PMNC}\}$ of the PMNC with respect to the 1–2–3 coordinate system turn out to be location dependant and can be determined by the following transformations:

$$[\bar{C}^{PMNC}] = [T]^{-T}[\bar{C}^{NC}][T]^{-1} \quad \text{and} \quad \{\bar{\alpha}^{PMNC}\} = [T]^{-T}\{\bar{\alpha}^{NC}\} \tag{10}$$

$$\text{where, } [T] = \begin{bmatrix} 1 & 0 & 0 & 0 & 0 & 0 \\ 0 & m^2 & n^2 & mn & 0 & 0 \\ 0 & n^2 & m^2 & -mn & 0 & 0 \\ 0 & -2mn & 2mn & m^2-n^2 & 0 & 0 \\ 0 & 0 & 0 & 0 & m & -n \\ 0 & 0 & 0 & 0 & n & m \end{bmatrix} \text{ with } m = \cos\theta \text{ and } n = \sin\theta$$

From Eq. (10) it is obvious that the effective thermoelastic properties at any point of the PMNC surrounding the carbon fiber with respect to the principle material coordinate (1–2–3) axes of the FFRC vary over an annular cross section of the PMNC phase of the RVE of the CFF. However, without loss of generality, it may be considered that the volume average of these effective thermoelastic properties over the volume of the PMNC can be treated as the constant effective elastic coefficient matrix $[C^{PMNC}]$ and the constant effective thermal expansion coefficient vector $\{\alpha^{PMNC}\}$ of the PMNC containing sinusoidally wavy CNTs surrounding the carbon fiber with respect to the 1–2–3 coordinate axes of the FFRC and are given by

$$[C^{PMNC}] = \frac{1}{\pi(R^2 - a^2)} \int_0^{2\pi} \int_a^R [\bar{C}^{PMNC}] \, r \, dr \, d\theta \quad \text{and}$$

$$\{\alpha^{PMNC}\} = \frac{1}{\pi(R^2 - a^2)} \int_0^{2\pi} \int_a^R \{\bar{\alpha}^{PMNC}\} \, r \, dr \, d\theta \tag{11}$$

Since the CFF is a composite in which the carbon fiber is the reinforcement and the matrix phase is the PMNC material, the MT model can be employed to estimate its effective elastic properties. Thus according to the MT model [31], the effective elastic coefficient matrix for the CFF is given by

$$[C^{CFF}] = [C^{PMNC}] + \bar{v}_f([C^f] - [C^{PMNC}])[\tilde{A}_2]\left[v_{PMNC}[I] + v_f[\tilde{A}_2]\right]^{-1} \qquad (12)$$

in which the matrix of the strain concentration factors are given by

$$[\tilde{A}_2] = \left[[I] + [S_2]([C^{PMNC}])^{-1}([C^f] - [C^{PMNC}])\right]^{-1} \qquad (13)$$

In Eqs. (12) and (13), $\bar{v}_f$ and $v_{PMNC}$ are the volume fractions of the carbon fiber and the PMNC material, respectively, with respect to the volume of the RVE of the CFF and the Eshelby tensor $[S_2]$ is computed based on the elastic properties of the PMNC matrix and the shape of the carbon fiber. It is worthwhile to note that the PMNC is transversely isotropic and consequently, the Eshelby tensor [35] corresponding to transversely isotropic material is utilized for computing the matrix $[S_2]$ while the inclusion is a circular cylindrical fiber. Using the effective elastic coefficient matrix $[C^{CFF}]$, the effective thermal expansion coefficient vector $\{\alpha^{CFF}\}$ for the CFF can be derived as follows [33]:

$$\{\alpha^{CFF}\} = \{\alpha^f\} + \left([C^{CFF}]^{-1} - [C^f]^{-1}\right)\left([C^f]^{-1} - [C^{PMNC}]^{-1}\right)^{-1}\left(\{\alpha^f\} - \{\alpha^{PMNC}\}\right) \qquad (14)$$

where $[C^f]$ and $\{\alpha^f\}$ are the elastic coefficient matrix and the thermal expansion coefficient vector of the carbon fiber, respectively. Finally, considering the CFF as the cylindrical inclusion embedded in the isotropic polymer matrix, the effective elastic properties $[C]$ of the FFRC can be derived by the MT method [31] as follows:

$$[C] = [C^p] + v_{CFF}([C^{CFF}] - [C^p])[\tilde{A}_3]\left[\bar{v}_p[I] + v_{CFF}[\tilde{A}_3]\right]^{-1} \qquad (15)$$

in which the matrix of the strain concentration factors are given by

$$[\tilde{A}_3] = \left[[I] + [S_1]([C^p])^{-1}([C^{CFF}] - [C^p])\right]^{-1} \qquad (16)$$

where $v_{CFF}$ and $\bar{v}_p$ are the volume fractions of the CFF and the polymer material, respectively, with respect to the volume of the RVE of the FFRC. Finally, the effective thermal expansion coefficient vector $\{\alpha\}$ for the FFRC can be derived as follows [33]:

$$\{\alpha\} = \{\alpha^{CFF}\} + ([C]^{-1} - [C^{CFF}]^{-1})([C^{CFF}]^{-1} - [C^p]^{-1})^{-1}(\{\alpha^{CFF}\} - \{\alpha^p\}) \qquad (17)$$

For the purpose of verifying the MT model, the effective thermoelastic properties predicted by this model should be compared with those predicted by different micromechanical model. Hence, another micromechanics model based on the method of cells (MOC) approach will be utilized here. However, for the sake brevity, the development of the MOC approach is not presented here and can be found in Ref. [36].

## 2.3 Effective thermal conductivities of FFRC

This Section deals with the procedures of employing two modeling approaches, namely, the MOC approach and the effective medium (EM) approach to predict the effective thermal conductivities of the FFRC. The various steps involved in the modeling of the thermal conductivities of the FFRC are outlined as follows:

- First, the effective thermal conductivities of the PMNC are to be determined by using either the MOC approach considering the perfect CNT/polymer matrix interface ($R_k = 0$) or the EM approach incorporating the CNT/polymer matrix interfacial thermal resistance ($R_k \neq 0$) where $R_k$ is the CNT/polymer matrix interfacial thermal resistance.
- Utilizing the thermal conductivities of the PMNC and the carbon fiber, the effective thermal conductivities of the CFF are to be determined by using the MOC approach.
- Finally, using the thermal conductivities of the CFF and the polymer matrix, the effective thermal conductivities of the FFRC can be estimated by employing the MOC approach.

### 2.3.1 Method of cells (MOC) approach

This Section presents the development of the MOC approach to estimate the effective thermal conductivities of the PMNC, the CFF and the FFRC. Assuming that CNTs are equivalent solid fibers [9, 10, 17–20], uniformly spaced in the polymer matrix and are aligned along the $x_3$–axis, the unwound PMNC can be viewed to be composed of cells forming doubly periodic arrays along the $x_1$– and $x_2$–directions. Figure 6 shows a repeating unit cell (RUC) with four subcells. Each rectangular subcell is labeled by β γ, with β and γ denoting the location of the subcell along the $x_1$– and $x_2$–directions, respectively. The subcell can be either a CNT or the polymer matrix. Let four local coordinate systems ($\bar{x}_1^{(\beta)}$, $\bar{x}_2^{(\gamma)}$ and $x_3$) be introduced, all of which have origins that are located at the centroid of each cell. In accordance with the MOC approach, the deviation of the temperature from a reference temperature $T_R$ (at which the material is stress free when its strain is zero), $\Delta\Theta^{(\beta\gamma)}$, is expanded in the following form:

$$\Delta\Theta^{(\beta\gamma)} = \Delta T + \bar{x}_1^{(\beta)}\xi_1^{(\beta\gamma)} + \bar{x}_2^{(\gamma)}\xi_2^{(\beta\gamma)} \tag{18}$$

where $\xi_1^{(\beta\gamma)}$ and $\xi_2^{(\beta\gamma)}$ characterize the linear dependence of the temperature on the local coordinates. The volume $(V_{\beta\gamma})$ of each subcell is

$$V_{\beta\gamma} = b_\beta h_\gamma l \tag{19}$$

where $b_\beta$, $h_\gamma$ and $l$ denote the width, the height and the length of the subcell, respectively, while the volume $(V)$ of the RUC is

$$V = bhl \tag{20}$$

The continuity conditions of the temperature at the interfaces of the subcells on an average basis lead to the following relations

$$h_1\xi_1^{(1\gamma)} + h_2\xi_1^{(2\gamma)} = (h_1 + h_2)\frac{\partial T}{\partial x_1}$$

$$b_1\xi_2^{(\beta 1)} + b_2\xi_2^{(\beta 2)} = (b_1 + b_2)\frac{\partial T}{\partial x_2} \tag{21}$$

For the average heat flux in the subcell:

$$\bar{q}_i^{(\beta\gamma)} = -K_i^{(\beta\gamma)}\frac{\partial T}{\partial x_3} \quad ; i = 1, 2, 3 \tag{22}$$

where $K_i^{(\beta\gamma)}$ denote the thermal conductivity coefficients of the subcells.

The average heat flux in the unwound PMNC material is determined from the following relation:

$$q_i = \frac{1}{V}\sum_{\beta,\gamma=1}^{2} V_{\beta\gamma}\bar{q}_i^{(\beta\gamma)} \tag{23}$$

The continuity conditions of the heat flux at the interfaces of the subcells yield

$$\bar{q}_1^{(1\gamma)} = \bar{q}_1^{(2\gamma)} \quad \text{and} \quad \bar{q}_2^{(\beta 1)} = \bar{q}_2^{(\beta 2)} \tag{24}$$

The average heat flux components are related to the temperature gradients by the effective thermal conductivity coefficients $(K_i^{nc})$:

$$\bar{q}_i = -K_i^{nc}\frac{\partial T}{\partial x_3} \tag{25}$$

By eliminating the microvariables $\xi_1^{(\beta\gamma)}$ and $\xi_2^{(\beta\gamma)}$, and using the continuity conditions, the effective thermal conductivities of the unidirectional unwound PMNC lamina are given by [37]

$$K_1^{nc} = \frac{K^p\{K^n[h(V_{11} + V_{21}) + h_2(V_{12} + V_{22})] + K^p h_1(V_{12} + V_{22})\}}{hbl(K^p h_1 + K^n h_2)},$$

$$K_2^{nc} = \frac{K^p\{K^n[b(V_{11} + V_{12}) + b_2(V_{21} + V_{22})] + K^p b_1(V_{21} + V_{22})\}}{hbl(K^p b_1 + K^n b_2)} \quad \text{and}$$

$$K_3^{nc} = \frac{K^n V_{11} + K^p (V_{12} + V_{21} + V_{22})}{hbl} \tag{26}$$

The effective thermal conductivities ($K_i^{NC}$) at any point in the unwound PMNC lamina where the CNT is inclined at an angle ϕ with the 3 (3')–axis can be derived in a straightforward manner by employing the appropriate transformation law as follows:

$$[K^{NC}] = [T_1]^{-T}[K^{nc}][T_1]^{-1} \quad \text{and} \quad [K^{NC}] = [T_2]^{-T}[K^{nc}][T_2]^{-1} \tag{27}$$

according as the wavy CNT is coplanar with the 2–3 (2'–3') or the 1–3 (1'–3') plane, respectively. The various matrices appeared in Eq. (27) are given by

$$[K^{nc}] = \begin{bmatrix} K_1^{nc} & 0 & 0 \\ 0 & K_2^{nc} & 0 \\ 0 & 0 & K_3^{nc} \end{bmatrix}, \quad [T_1] = \begin{bmatrix} 1 & 0 & 0 \\ 0 & \cos\phi & \sin\phi \\ 0 & -\sin\phi & \cos\phi \end{bmatrix} \quad \text{and} \quad [T_2] = \begin{bmatrix} \cos\phi & 0 & \sin\phi \\ 0 & 1 & 0 \\ -\sin\phi & 0 & \cos\phi \end{bmatrix}$$

Following the procedure for deriving the effective elastic coefficient matrix $[C^{PMNC}]$, the effective thermal conductivity matrix $[K^{PMNC}]$ can be obtained as follows:

$$[\bar{K}^{NC}] = \frac{1}{L_n} \int_0^{L_n} [K^{NC}] \, dy \tag{28}$$

$$[\bar{K}^{PMNC}] = \begin{bmatrix} 1 & 0 & 0 \\ 0 & \cos\theta & \sin\theta \\ 0 & -\sin\theta & \cos\theta \end{bmatrix}^{-T} \begin{bmatrix} K_1^{nc} & 0 & 0 \\ 0 & K_2^{nc} & 0 \\ 0 & 0 & K_3^{nc} \end{bmatrix} \begin{bmatrix} 1 & 0 & 0 \\ 0 & \cos\theta & \sin\theta \\ 0 & -\sin\theta & \cos\theta \end{bmatrix}^{-1} \tag{29}$$

$$[K^{PMNC}] = \frac{1}{\pi(R^2 - a^2)} \int_0^{2\pi} \int_a^R [\bar{K}^{PMNC}] \, r \, dr \, d\theta \tag{30}$$

It is worthwhile to note that the thermal conductivity matrix of the homogenized PMNC $[K^{PMNC}]$ is transversely isotropic and its axis of transverse isotropy is the 1- or $x_1$– axis. In order to model the CFF by the MOC approach, the CFF is considered to be composed of cells periodically arranged along the $x_2$– and $x_3$–directions while each cell consists of βγ number of subcells. In this case, each RUC represents the CFF and the subcell is composed of either the carbon fiber or the PMNC. Finally, the MOC approach for the CFF can be augmented in a straightforward manner to estimate the effective thermal conductivities of the FFRC in which the polymer is the matrix material and the CFF is the reinforcement along the 1- or $x_1$- direction.

*2.3.1 Effective medium (EM) approach*

This Section presents the Maxwell Garnett type EM approach to estimate the effective thermal conductivities of the PMNC incorporating the CNT/polymer matrix interfacial thermal resistance. Assuming CNTs as solid fibers [9, 10, 17–20], the EM approach by Nan et al. [38] can be augmented to predict the effective thermal conductivities ($K_i^{nc}$) of the unwound PMNC material with straight CNTs and are given by

$$K_1^{nc} = K_2^{nc} = K^p \frac{K^n(1+\alpha) + K^p + v_n[K^n(1-\alpha) - K^p]}{K^n(1+\alpha) + K^p - v_n[K^n(1-\alpha) - K^p]} \quad \text{and} \quad K_3^{nc} = v_n K^n + v_p K^p \quad (31)$$

In Eq. (31), a dimensionless parameter $\alpha = 2a_k/d_n$ in which the interfacial thermal property is concentrated on a surface of zero thickness and characterized by Kaptiza radius, $a_k = R_k K^p$, where $d_n$ represents the diameter of the CNT. Once $[K^{nc}]$ is computed, Eqs. (27) – (30) are used to estimate the effective thermal conductivities of the PMNC material surrounding the carbon fiber.

## 3. NOVEL FUZZY CARBON FIBER HEAT EXCHANGER

Heat dissipating systems such as microelectronics, transportation, heat exchangers etc. require an efficient heat removal capacity to avoid possible damage due to thermal stresses. The conventional method for increasing heat dissipation is to increase the surface area available for exchanging heat with a heat transfer fluid. However, this approach requires an undesirable increase in the thermal management system's size. There is therefore an urgent need for the development of novel advanced structures with better heat transfer performance. Thus the current status of progress in research on CNT-reinforced composites brings to light that the three-phase hybrid CNT-reinforced composite can be the promising candidate material for achieving better thermal management benefits from the exceptionally conductive CNTs. To fulfill the demand of better heat dissipating systems, a novel fuzzy carbon fiber heat exchanger (FFHE) composed of highly conductive CNTs is also studied in this study. A novel constructional feature of the FFHE is that CNTs are radially grown on the outer circumferential surface of the hollow cylindrical carbon fiber (HCF) heat exchanger as shown in Fig. 7. Analytical models based on the MOC and the EM approaches derived in Section 2.3 have been utilized for predicting the effective thermal conductivities of this proposed FFHE. In the present study, an attempt has also been made to investigate the effects of waviness of CNTs and CNT/polymer matrix interfacial thermal resistance on the heat transfer performance of the FFHE.

## 4. RESULTS AND DISCUSSION

In this Section, the predictions by the micromechanics models developed herein are first compared with those of the existing experimental and numerical results. Subsequently, the effective properties of the FFRC and the FFHE have been determined.

### 4.1 Comparisons with experimental and numerical results

Kulkarni et al. [11] experimentally and numerically investigated the elastic response of the nano-reinforced laminated composite (NRLC). The NRLC is made of the CNT-reinforced polymer nanocomposite and the carbon fiber. The geometry of the NRLC as shown in Fig. 8 is similar to that of the CFF shown in Fig. 3 if straight CNTs are considered. Thus to confirm the modeling of the CFF in the present study, the comparisons have been made between the results predicted by Kulkarni et al. [11] for the NRLC with those of the results predicted by the MT and the MOC models for the CFF with straight CNTs. It may be observed from Table 1 that the predicted value of the transverse Young's modulus ($E_x$) of the CFF computed by the MT and the MOC models match closely with that of the experimental value predicted by Kulkarni et al. [11]. The experimental value of $E_x$ is lower than the theoretical prediction and this may be attributed to the fact that CNTs are not perfectly radially grown and straight, and hence the radial stiffening of the NRLC decreases [11]. Further possible reasons for the disparity between the analytical and the experimental results include the lattice defects within CNTs and the formation of voids in CNT-reinforced composite [12]. It may also be noted that the value of $E_x$ predicted by the MT and the MOC models utilized herein is much closer to the experimental value than that of the numerical value predicted by Kulkarni et al. [11]. This is attributed to the fact that the appropriate transformation and homogenization procedures given by Eqs. (10) and (11) have been employed in the present study whereas Kulkarni et al. [11] did not consider such transformation and homogenization procedures in their numerical modeling. These comparisons are significant since the prediction of the transverse Young's modulus of the CFF provides critical check for the validity of the MT and the MOC models. Thus it can be inferred from the comparisons shown in Table 1 that the MT and the MOC models can be reasonably applied to predict the elastic properties of the FFRC and its phases.

Thermal conductivities predictions by the MOC and the EM approaches derived in the present study are first compared with the experimental results by Marconnet et al. [12]. Marconnet et al. [12] fabricated the aligned CNT-polymer nanocomposites consisting of CNTs arrays infiltrated with an aerospace-grade thermoset epoxy. In their study, the axial and the transverse thermal conductivities of the aligned CNT-polymer nanocomposites are found to be in good agreement with those of the values estimated by using the EM approach. The comparisons of the axial ($K_A$) and the transverse ($K_T$) thermal conductivities of the aligned CNT-polymer nanocomposites estimated by the MOC and the EM approaches with those of the experimental results are illustrated in Figs. 9 (a) and (b), respectively. In these figures, dotted blue line represents best fits obtained from the EM approach for the experimental results considering an alignment factor (AF) of CNTs as 0.77 [12]. Figure 9 (a) reveals that the effective values of $K_A$ predicted by the MOC and the EM approaches overestimate the experimental values of $K_A$ by ~21% and ~23% when the values of the CNT volume fractions are 0.07 and 0.16, respectively. On the other hand, the MOC and the EM approaches underestimate the values of $K_T$ by 40% and 58% when the values of the CNT volume fractions are 0.07 and 0.16, respectively. These differences between the results are attributed to the fact that the perfect alignments of CNTs (i.e., AF = 1) are considered while computing the results by the MOC and the EM approaches whereas the value of the AF is 0.77 in Ref. [12]. Other possible reasons for the disparity between the analytical and the experimental results include the CNT/matrix interfacial thermal resistance, lattice defects within CNTs and modification of the phonon conduction within CNTs due to interactions with the matrix [9, 10, 12]. These comparisons also reveal that the thermal conductivities predicted by the EM approach agree with those predicted by the MOC approach. Thus it can be inferred from the comparisons shown in Figs. 9 (a) and (b) that the MOC and the EM approaches can be reasonably applied to predict the thermal conductivities of the FFRC.

## 4.2 Analytical modeling results

To present the numerical results, the coefficients of thermal expansion (CTEs) and the thermal conductivities of the carbon fiber, armchair (10, 10) CNT and the polymer matrix are considered to be temperature dependent and are taken from Ref [7, 39–43]. It is obvious that the constructional feature of the FFRC imposes a constraint on the maximum value of the CNT volume fraction. The maximum value of the CNT volume fraction in the FFRC can be determined based on the surface to surface distance at the roots of two adjacent CNTs as 1.7 nm [17], the CNT diameter ($d_n$), the running length of the sinusoidally wavy CNT ($L_{nr}$) and the volume fraction of the carbon fiber ($v_f$) as follows [17]:

$$(V_{CNT})_{max} = \frac{\pi d_n^2 L_{nr} v_f}{d(d_n + 1.7)^2} \tag{32}$$

The effect of waviness of CNTs on the effective thermomechanical properties of the FFRC is investigated when the wavy CNTs are coplanar with either of the two mutually orthogonal planes. For such investigation, the values of the volume fraction of the carbon fiber ($v_f$) in the FFRC and the maximum amplitude of the CNT wave (A) are considered as 0.5 and $100d_n$ nm (i.e., 136 nm), respectively. If the value of $v_f$ is 0.5 then the diameter of the CFF (2R) turns out to be 13.4677 μm and the corresponding value of the straight CNT length ($L_n$) in the CFF is 1.734 μm. The degree of waviness of the CNT is defined by the waviness factor ($A/L_n$). It should be noted that for the straight CNT, the value of $A/L_n$ is zero.

Unless otherwise mentioned, the effective thermomechanical properties of the PMNC are computed by employing the MOC approach considering the perfect CNT/polymer matrix interface without any interfacial thermal resistance. Subsequently, the estimated effective thermomechanical properties of the PMNC are used to compute the effective thermomechanical properties of the CFF. However, for the sake of brevity, the effective thermomechanical properties of the PMNC and the CFF are not presented here. The variations of the amplitudes of CNT waves are considered for the two particular values of ω (i.e., $12\pi/L_n$ and $24\pi/L_n$). Figures 10 and 11 illustrate the variations of the effective elastic coefficients $C_{11}$ and $C_{12}$ of the FFRC with the waviness factor ($A/L_n$), respectively. It may be observed from Fig. 10 that the effective values of $C_{11}$ of the FFRC are not affected by the variations of the amplitude of the wavy CNTs in the 2–3 plane. When the wavy CNTs are coplanar with the 1–3 plane, the increase in the values of $A/L_n$ and $\omega$ significantly enhances the value of $C_{11}$. Figure 11 reveals that the waviness of CNTs causes significant increase in the value of $C_{12}$ when the wavy CNTs are coplanar with the 1–3 plane. Since the FFRC is transversely isotropic material, the values of $C_{13}$ are found to be identical to those of the values of $C_{12}$. Figure 12 reveals that the increase in the value of $A/L_n$ decreases the value of $C_{22}$ when the CNT waviness is coplanar with the 1–3 plane whereas the value of $C_{22}$ enhances for the higher values of $A/L_n$ and $\omega$ when the CNT waviness is coplanar with the 2–3 plane. Although not presented here, the similar trends of results are obtained for the effective elastic coefficients $C_{23}$, $C_{23}$ and $C_{44}$. It may be noted from Figs. 10–12 that if the wavy CNTs are coplanar with the 1–3 plane then the axial elastic coefficients of the FFRC are significantly improved over their values with the straight CNTs ($\omega = 0$) for the higher values of $A/L_n$ and $\omega$. When the wavy CNTs are coplanar with the longitudinal plane (1–3 or 1'–3' plane) of the carbon fiber as shown in Fig. 3 (b), the amplitudes of the CNT waves becomes parallel to the 1–axis and this results into the aligning of the projections of parts of CNTs lengths with the 1–axis leading to the axial stiffening of the PMNC. The more is the value of $\omega$, the more will be such projections and hence the effective axial elastic coefficients ($C_{11}$, $C_{12}$, $C_{13}$, $C_{55}$ and $C_{66}$) of the FFRC increases with the increase in the value of $\omega$. On the other hand, if the wavy CNTs are coplanar with the transverse plane (2–3 or 2'–3' plane) of the carbon fiber then the transverse elastic coefficients ($C_{22}$, $C_{23}$, $C_{33}$ and $C_{44}$) of the FFRC are improved over their values with the straight CNTs ($\omega = 0$) and the reverse is true when the when the wavy CNTs are coplanar with the 1–3 (1'–3') plane.

Figures 13 and 14 illustrate the variations of the axial CTE ($\alpha_1$) and the transverse CTE ($\alpha_2$) of the FFRC with the waviness factor, respectively. It may be observed from Fig. 13 that the values of $\alpha_1$ of the FFRC are not affected by the variations of the amplitude of the wavy CNTs in the 2–3 plane whereas the values of $\alpha_1$ initially increases and then significantly decreases for the higher values of $A/L_n$ and $\omega$ when the CNT waviness is coplanar with the 1–3 plane. It is also important to note from Fig. 13 that the effective value of $\alpha_1$ is zero for the values of $A/L_n$ and $\omega$ as 0.048 and $24\pi/L_n$, respectively, when the wavy CNTs are coplanar with the 1–3 plane. Figure 14 reveals that the waviness of CNTs improves the effective values $\alpha_2$ of the FFRC when the wavy CNTs are coplanar with the 2–3 plane compared to that of the FFRC with the straight CNTs ($\omega = 0$). Although not presented here, the computed effective values of $\alpha_3$ are found to match identically with those of $\alpha_2$ corroborating the fact that the FFRC is transversely isotropic material. For the value of $A/L_n \geq 0.035$, the effective CTEs ($\alpha_1$, $\alpha_2$ and $\alpha_3$) of the FFRC with the wavy CNTs being coplanar with the 1–3 plane start to decrease. This is attributed to the fact that the negative axial and transverse CTEs ($\alpha_1^n$ and $\alpha_3^n$) of the radially grown wavy CNTs on the circumferential surfaces of the carbon fibers significantly suppress the positive CTE ($\alpha^p = 66 \times 10^{-6} K^{-1}$) of the polymer matrix which eventually lowers the effective values of $\alpha_1$ of the FFRC and this effect becomes more pronounced for the higher values $A/L_n$ and $\omega$ because the CNT volume fraction in the FFRC increases with the values of $A/L_n$ and $\omega$.

Figure 15 illustrates the variation of the axial thermal conductivity ($K_1$) of the FFRC with the temperature. Figure 15 reveals that if the variations of the amplitude of the wavy CNTs are in the 1–3 plane, the effective values of $K_1$ are significantly improved over those of the FFRC containing either the wavy CNTs being coplanar with the 2–3 plane or the straight CNTs. When compared with the results of the base composite (i.e., $V_{CNT} = 0$), almost 383% and 120% enhancements are occurred in the values of $K_1$ if the values of the temperature are 300 K and 400 K, respectively, and the waviness CNTs is coplanar with the 1–3 plane with $V_{CNT} = 0.2131$. It may also be observed from Fig. 15 that the effective values of $K_1$ decrease with the increase in the temperature when the wavy CNTs are coplanar with the 1–3 plane. This is due to the fact that the thermal conductivity ($K^n$) of the armchair CNT (10, 10) decreases with the increase in the temperature which eventually lowers the effective value of $K_1$ of the FFRC. It may also be importantly observed from Fig. 15 that the effective values of $K_1$ of the FFRC containing either the wavy CNTs being coplanar with the 2–3 plane or the straight CNTs ($\omega = 0$) are not improved compared to those of the base composite ($V_{CNT} = 0$). Figure 16 depicts that the effective transverse thermal conductivities ($K_2$) of the FFRC are significantly improved over those of the base composite (i.e., $V_{CNT} = 0$) irrespective of the values of the CNT wave frequency and the planer orientations of the wavy CNTs. When compared with the results of the base composite (i.e., $V_{CNT} = 0$), nearly 640% enhancement is occurred in the values of $K_2$ for the temperature range 250K – 400 K if the CNTs (straight or wavy) are present on the circumferential surfaces of the carbon fiber reinforcements with minimum $V_{CNT} = 0.0538$ which is corresponding to the value of $\omega = 0$ (i.e., straight CNTs). This is attributed to the fact that the highly conductive CNTs being in the transverse plane of the carbon fiber leads to the increase in the transverse thermal conductivities of the PMNC along its radial direction which eventually enhance the transverse thermal conductivities of the FFRC.

So far the effective thermal conductivities of the FFRC have been estimated considering the perfect CNT/polymer matrix interface without any interfacial thermal resistance ($R_k = 0$). However, the CNT/polymer matrix interfacial thermal resistance may affect the heat transfer characteristics of the FFRC. Researchers reported that the magnitude of $R_k$ between nanoparticles/CNTs and different matrices ranges from $0.77 \times 10^{-8}$ m$^2$K/W to $20 \times 10^{-8}$ m$^2$K/W m$^2$K/W [10, 44]. To analyze the effect of the CNT/polymer matrix interfacial thermal resistance on the effective thermal conductivities of the FFRC, the EM approach incorporating such CNT/polymer matrix interfacial thermal resistance has been utilized and the values of $R_k$ are varied from 0 to $20 \times 10^{-8}$ m$^2$K/W. Figure 17 illustrates the variation of the effective values of $K_1$ of the FFRC incorporating the CNT/polymer matrix interfacial thermal resistance when the waviness of CNTs is coplanar with the 1–3 plane. This figure reveals that the values of $K_1$ are independent of the values of $R_k$. Although not presented here, the values of $K_2$ and $K_3$ are also found to be independent of the values of $R_k$.

The effect of waviness of CNTs on the effective thermal conductivities of the FFHE is investigated when the wavy CNTs are coplanar with either of the two mutually orthogonal planes with $\omega = 24\pi/L_n$. For comparison purpose, the values of the inner ($d_i$) and the outer ($d_o$) diameters of the bare HCF (without CNTs) heat exchanger and the FFHE are kept constant to 50 μm and 100 μm, respectively. The maximum value of the CNT volume fraction in the FFHE can be determined based on the surface to surface distance at the roots of two adjacent CNTs as 1.7 nm [17], the CNT diameter ($d_n$), the running length of the sinusoidally wavy CNT ($L_{nr}$) and the outer diameter of the HCF (d) as follows:

$$(V_{CNT})_{max} = \frac{V^{CNT}}{V^{FFHE}} = \frac{\pi d_n^2 d L_{nr}}{(d_o^2 - d_i^2)(d_n + 1.7)^2} \tag{33}$$

The derivation of Eq. (33) has been presented in Appendix A. The value of the diameter of the HCF (d) in the FFHE is considered as 60 μm to evaluate the numerical results. Figures 18 and 19 illustrate the variations of the axial ($K_1$) and the transverse ($K_2$) thermal conductivities of the FFHE, respectively. It may be observed from Fig. 18 that the effective values of $K_1$ of the FFHE are not affected by the variations of the amplitude of the wavy CNTs in the 2–3 plane. On the other hand, the effective values of $K_1$ are significantly improved with the increase in the values of $A/L_n$ if the wavy CNTs are coplanar with the 1–3 plane. Figure 19 depicts that the effective values of $K_2$ are slightly improved for the higher values of $A/L_n$ when the wavy CNTs are coplanar with the 2–3 plane. Although not presented here, the computed effective values of $K_3$ are found to match identically with those of $K_2$ corroborating the fact that the FFHE is transversely isotropic about the 1-axis and the effective thermal conductivities of the FFHE are independent of the values of $R_k$.

## 5. CONCLUSIONS

Hierarchical modeling of the multifunctional CNT-reinforced hybrid composites such as FFRC and FFHE have been carried out in the present study. The constructional feature of such multifunctional CNT-reinforced hybrid composites is that the amplitudes of the sinusoidally wavy CNTs radially grown on the circumferential surfaces of the carbon fibers are either parallel or transverse to the axis of the carbon fibers. Thermomechanical properties of the FFRC and the FFHE are estimated by employing the MT model, MOC approach and the EM approach. The following main inferences are drawn from the present study:

1. If the plane of the radially grown wavy CNTs is coplanar with the longitudinal plane of the carbon fiber then the axial effective elastic and thermoelastic coefficients of the FFRC are significantly improved over those of the FFRC with either the straight CNTs ($\omega = 0$) and or the wavy CNTs being coplanar with the transverse plane of the carbon fiber. When the CNT waviness is coplanar with the transverse plane of the carbon fiber then the transverse effective elastic and thermoelastic coefficients of the FFRC are improved.

2. If the amplitudes of the sinusoidally wavy CNTs are coplanar with the axis of the carbon fiber or the HCF, the effective thermal conductivities of the FFRC and FFHE are significantly improved over those of the base composites (i.e., without CNTs). On the other hand, the effective transverse thermal conductivities of the FFRC and the FFHE are significantly enhanced over those of the base composites irrespective of the values of the CNT wave frequency and the planer orientations of the wavy CNTs.

3. The CNT/polymer matrix interfacial thermal resistance does not affect the effective thermal conductivities of the FFRC and the FFHE.

4. The effective thermomechanical properties of the FFRC and the FFHE estimated by the analytical micromechanics models based on the MT method, the MOM approach and the EM approach are in excellent agreement with those of the existing experimental and numerical results. Hence, for predicting all effective properties of the advanced hierarchical multifunctional CNT-reinforced hybrid composite, one can adopt the analytical micromechanics model.

Since the effective thermomechanical properties of the multifunctional CNT-reinforced hybrid composite containing sinusoidally wavy CNTs are significantly enhanced and can be altered as per requirement, the multifunctional CNT-reinforced hybrid composite may be used for developing high-performance light weight structures or heat exchangers which require stringent constraint on the dimensional stability with enhanced thermal management capability.

# APPENDIX A

The maximum value of the CNT volume fraction in the FFHE can be determined as follows:

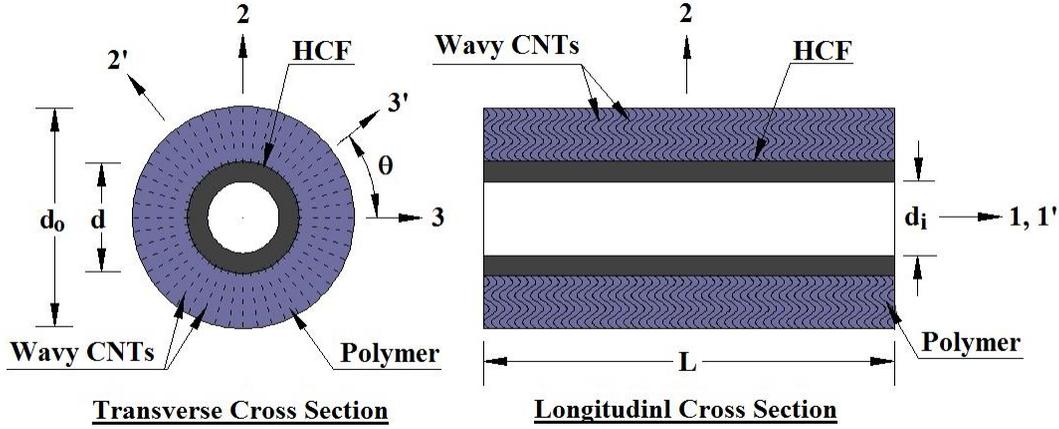

*Fig. A1 Transverse and longitudinal cross sections of the FFHE containing wavy CNTs being coplanar with the 1–3 plane*

Referring to Fig. A1, the volumes of the HCF ($V^f$), the PMNC ($V^{PMNC}$) and the FFHE ($V^{FFHE}$) are given by

$$V^f = \frac{\pi}{4}(d^2 - d_i^2)L \qquad \text{A (1)}$$

$$V^{PMNC} = \frac{\pi}{4}(d_o^2 - d^2)L \qquad \text{A (2)}$$

$$V^{FFHE} = \frac{\pi}{4}(d_o^2 - d_i^2)L \qquad \text{A (3)}$$

Using Eqs. A (1) and (3), the volume fraction of the HCF ($v_f$) in the FFHE can be determined as

$$v_f = \frac{V^f}{V^{FFHE}} = \frac{(d^2 - d_i^2)}{(d_o^2 - d_i^2)} \qquad \text{A (4)}$$

The maximum number of radially grown aligned CNTs $(N_{CNT})_{max}$ on the outer circumferential surface of the HCF is given by

$$(N_{CNT})_{max} = \frac{\pi d L}{(d_n + 1.7)^2} \qquad \text{A (5)}$$

Therefore, the volume of the CNTs ($V^{CNT}$) is

$$V^{CNT} = \frac{\pi}{4} d_n^2 L_{nr} (N_{CNT})_{max} \qquad \text{A (6)}$$

Thus the maximum volume fraction of the CNTs $(V_{CNT})_{max}$ with respect to the volume of the FFHE can be determined as

$$(V_{CNT})_{max} = \frac{V^{CNT}}{V^{FFHE}} = \frac{\pi d_n^2 d L_{nr}}{(d_o^2 - d_i^2)(d_n + 1.7)^2} \qquad \text{A (7)}$$

The maximum volume fraction of the CNTs with respect to the volume of the PMNC $(v_n)_{max}$ can be determined in terms of $(V_{CNT})_{max}$ as follows:

$$(v_n)_{max} = \frac{V^{CNT}}{V^{PMNC}} = \frac{\pi d_n^2 d L_{nr}}{(d_o^2 - d^2)(d_n + 1.7)^2} \qquad \text{A (8)}$$

# REFERENCES

1. Iijima, S. (1991): "Helical microtubules of graphitic carbon". *Nature*, Vol. 354, pp. 56–58.
2. Treacy, M. M. J.; Ebbesen, T. W. and Gibson, J. M. (1996): "Exceptionally high Young's modulus observed for individual carbon nanotubes". *Nature*, Vol. 381, pp. 678–680.
3. Li, C. and Chou, T. W. (2003): "A Structural mechanics approach for the analysis of carbon nanotubes". *Int. J. Solids Struct.*, Vol. 40, pp. 2487–2499.
4. Shen, L. and Li, J. (2004): "Transversely isotropic elastic properties of single-walled carbon nanotubes". *Phys. Rev. B*, Vol. 69, p. 045414.
5. Tsai, J. L.; Tzeng, S. H. and Chiu, Y. T. (2010): "Characterizing elastic properties of carbon nanotube/polyimide nanocomposites using multi-scale simulation". *Compos. Part B*, Vol. 41, pp. 106–115.
6. Hone, J.; Whitney, C.; Piskoti, C. and Zettl, A. (1999): "Thermal conductivity of single-walled carbon nanotubes". *Phys. Rev. B*, Vol. 59, No. 4, pp. 2514–2516.
7. Berber, S.; Kwon, Y. K. and Tomanek, D. (2000): "Unusually high thermal conductivity of carbon nanotubes". *Phys. Rev. Lett.*, Vol. 84, No. 20, pp. 4613–4616.
8. Yu, C.; Shi, L.; Yao, Z.; Li, D. and Majumdar, A. (2005): "Thermal conductance and thermopower of an individual single-wall carbon nanotube". *Nano Lett.*, Vol. 5, No. 9, pp. 1842–1846.
9. Biercuk, M. J.; Llaguno, M. C.; Radosavljevic, M.; Hyun, J. K.; Johnson, A. T. and Fischer, J. E. (2002): "Carbon nanotube composites for thermal management". *Appl. Phys. Lett.*, Vol. 80, No. 15, pp. 2767–2769.
10. Bryning, M. B.; Milkie, D. E.; Islam, M. F.; Kikkawa, J. M. and Yodh, A. G. (2005): "Thermal conductivity and interfacial resistance in single-wall carbon nanotube epoxy composites". *Appl. Phys. Lett.*, Vol. 87, No. 16, p. 161909.
11. Kulkarni, M.; Carnahan, D.; Kulkarni, K.; Qian, D. and Abot, J. L. (2010): "Elastic response of a carbon nanotube fiber reinforced polymeric composite: A numerical and experimental study". *Compos. Part B*, Vol. 41, No. 5, pp. 414–421.
12. Marconnet, A. M.; Yamamoto, N.; Panzer, M. A.; Wardle, B. L. and Goodson, K. E. (2011): "Thermal conduction in aligned carbon nanotube-polymer nanocomposites with high packing density". *ACS Nano.*, Vol. 5, No. 6, pp. 4818–4825.
13. Thostenson, E. T.; Ren, Z. and Chou, T. W. (2001): "Advances in the science and technology

of carbon nanotubes and their composites: a review". *Compos. Sci. Technol.*, Vol. 61, No. 13, pp. 1899–1912.

14. Wernik, J. M. and Meguid, S. A. (2010): "Recent developments in multifunctional nanocomposites using carbon nanotubes". *ASME Appl. Mech. Rev.*, Vol. 63, p. 050801.

15. Veedu, V. P.; Cao, A.; Li, X.; Ma, K.; Soldano, C.; Kar, S.; Ajayan, P. M. and Ghasemi-Nejhad, M. N. (2006): "Multifunctional composites using reinforced laminae with carbon-nanotube forests". *Nature Mater.*, Vol. 5, pp. 457–462.

16. Garcia, E. J.; Wardle, B. L.; Hart, A. J. and Yamamoto, N. (2008): "Fabrication and multifunctional properties of a hybrid laminate with aligned carbon nanotubes grown *in situ*". *Compos. Sci. Technol.*, Vol. 68, No. 9, pp. 2034–2041.

17. Kundalwal, S. I. and Ray, M. C. (2011): "Micromechanical analysis of fuzzy fiber reinforced composites". *Int. J. Mech. Mater. Design*, Vol. 7, No. 2, pp. 149–166.

18. Kundalwal, S. I. and Ray, M. C. (2013): "Effect of carbon nanotube waviness on the elastic properties of the fuzzy fiber reinforced composites". *ASME J. Appl. Mech.*, Vol. 80, p. 021010.

19. Ray, M. C and Kundalwal, S. I. (2013): "A thermomechanical shear lag analysis of short fuzzy fiber reinforced composite containing wavy carbon nanotubes". *European J Mech–A/Solids.*, Vol. 44, pp. 41–60.

20. Kundalwal, S. I, Suresh Kumar, R and Ray, M. C. (2013): "Smart damping of laminated fuzzy fiber reinforced composite shells using 1-3 piezoelectric composites". *Smart Mater Struct.*, Vol. 22, No. 10, pp. 105001/1–16.

21. Kundalwal, S. I and Ray, M. C. (2014): "Shear lag analysis of a novel short fuzzy fiber-reinforced composite". *Acta Mech.*, Vol. 225, No. 9, pp. 2621–2643.

22. Ray, M. C and Kundalwal, S. I. (2014): "Effect of carbon nanotube waviness on the load transfer characteristic of the short fuzzy fiber-reinforced composites". *ASCE J Nanomech Micromech.*, Vol. 4, No. 2, p. A4013009.

23. Kundalwal, S. I and Ray, M. C. (2014): "Improved thermoelastic coefficients of a novel short fuzzy fiber-reinforced composite with wavy carbon nanotubes". *J Mech Mater Struct.*, Vol. 9, No. 1, pp. 1–25.

24. Kundalwal, S. I.; Suresh Kumar, R. and Ray, M. C. (2014): "Effective thermal conductivities of a novel fuzzy fiber reinforced composite containing wavy carbon nanotubes". *ASME J*

*Heat Trans.*, Vol. 137, No. 1, p. 012401.

25. Kundalwal, S. I. and Meguid, S. A. (2015): "Effect of carbon nanotube waviness on the active damping of laminated hybrid composite shells". *Acta Mech.*, Vol. 226, No. 6, pp. 2035–2052.

26. Kundalwal, S. I. and Ray, M. C. (2016): "Smart damping of fuzzy fiber reinforced composite plates using 1-3 piezoelectric composites". *J Vibration Control*, Vol. 22, No. 6, pp. 1526–1546.

27. Kundalwal, S. I. Kumar, S. (2016): "Multiscale modeling of microscale fiber reinforced composites with nano-engineered interphases". *Mech Mater.*, Vol. 102, pp. 117–131.

28. Fisher, F. T.; Bradshaw, R. D. and Brinson, L. C. (2002): "Effects of nanotube waviness on the modulus of nanotube-reinforced polymers". *Appl. Phys. Lett.*, Vol. 80, No. 24, pp. 4647–4649.

29. Berhan, L.; Yi, Y.B. and Sastry, A. M. (2004): "Effect of nanorope waviness on the effective moduli of nanotube sheets". *J. Appl. Phys.*, Vol. 95, No. 9, pp.5027–5034.

30. Anumandla, V. and Gibson, R. F. (2006): "A comprehensive closed form micromechanics model for estimating the elastic modulus of nanotube-reinforced composites". *Compos. Part A*, Vol. 37, No. 12, pp. 2178–2185.

31. Mori, T. and Tanaka, K. (1973): "Average stress in matrix and average elastic energy of materials with misfitting inclusions," *Acta Metall.*, Vol. 21, No. 5, pp. 571–574.

32. Qui, Y. P. and Weng, G. J. (1990): "On the application of Mori-Tanaka's theory involving transversely isotropic spheroidal inclusions". *Int. J. Eng. Sci.*, Vol. 28, No. 11, pp. 1121–1137.

33. Laws, N.: (1973): "On the thermostatics of composite materials," *J. Mech. Phys. Solids.*, Vol. 21, No. 1, pp. 9–17.

34. Hsiao, H. M. and Daniel, I. M. (1996): "Elastic properties of composites with fiber waviness". *Compos. Part A*, Vol. 27, No. 10, pp. 931–941.

35. Li, J. Y. and Dunn, M. L. (1998): "Anisotropic coupled-field inclusion and inhomogeneity problems". *Philos. Mag. A*, Vol. 77, No. 5, pp. 1341–1350.

36. Kundalwal, S. I. and Ray, M. C. (2013): "Thermoelastic properties of a novel fuzzy fiber-reinforced composite". *ASME J. Appl. Mech.*, Vol. 80, p. 061011.

37. Aboudi, J.; Arnold, S. M. and Bednarcyk, B. A. (2012): *"Micromechanics of Composite*

*Materials: A Generalized Multiscale Analysis Approach"*. Oxford OX5: Butterworth-Heinemann Ltd.

38. Nan, C. W.; Birringer, R.; Clarke, D. R. and Gleiter, H. (1997): "Effective thermal conductivity of particulate composites with interfacial thermal resistance". *J. Appl. Phys.*, Vol. 81, No. 10, pp. 6692–6699.

39. Kwon, Y. K.; Berber, S. and Tomanek, D. (2004): "Thermal contraction of carbon fullerenes and nanotubes". *Phys. Rev. Lett.*, Vol. 92, No. 1, p. 015901.

40. Honjo, K. (2007): "Thermal stresses and effective properties calculated for fiber composites using actual cylindrically-anisotropic properties of interfacial carbon coating". *Carbon*, Vol. 45, No. 4, pp. 865–872.

41. Villeneuve, J. F.; Naslain, R.; Fourmeaux, R. and Sevely, J. (1993): "Longitudinal/radial thermal expansion and poisson ratio of some ceramic fibers as measured by transmission electron microscopy". *Compos. Sci. Technol.*, Vol. 49, No. 1, pp. 89–103.

42. Wang, J. L.; Gu, M.; Ma, W. G.; Zhang, X. and Song, Y. (2008): "Temperature dependence of the thermal conductivity of individual pitch-derived carbon fibers". *New Carbon Mater.*, Vol. 23, pp. 259–63.

43. Reese, W. (1996): "Low-temperature thermal conductivity of amorphous polymers: polystyrene and polymethylmethacrylate". *J. Appl. Phys.*, Vol. 37, No. 2, pp. 864–868.

44. Wilson, O. M.; Hu, X.; Cahill, D. G. and Braun, P. V. (2002): "Colloidal metal particles as probes of nanoscale thermal transport in fluids". *Phys. Rev. B,* Vol. 66, No. 22, p. 224301.

**Table 1: Comparisons of the effective engineering constants of the NRLC with those of the CFF**

|  | NRLC* (2% CNT and 41% IM7 | | MT | MOC |
|---|---|---|---|---|
|  | Numerical [11] | Experimental [11] |  |  |
| $E_x$ | 13.93 | 10.02 | 11.50 | 11.91 |
| $\nu_{xy}$ | 0.34 | – | 0.38 | 0.38 |
| $\nu_{zx}$ | 0.16 | – | 0.17 | 0.182 |

where $E_x$ is the transverse Young's modulus of the NRLC; $\nu_{zx}$ and $\nu_{xy}$ are the axial Poisson's ratio and the transverse Poisson's ratio of the NRLC, respectively.

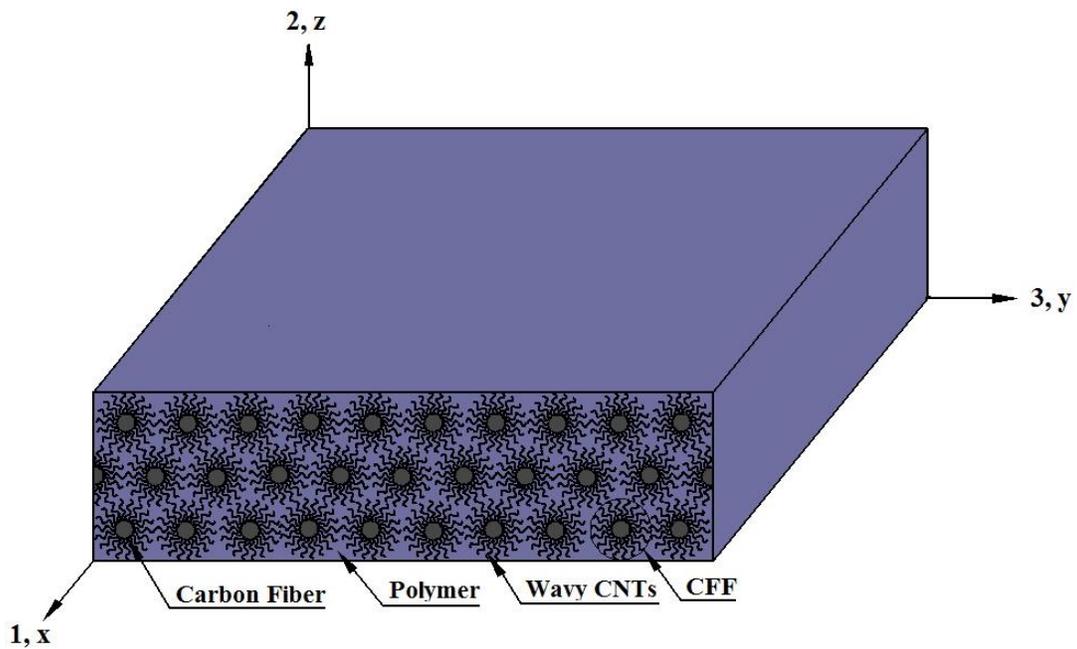

*Fig. 1 Schematic diagram of a lamina made of the FFRC containing wavy CNTs*

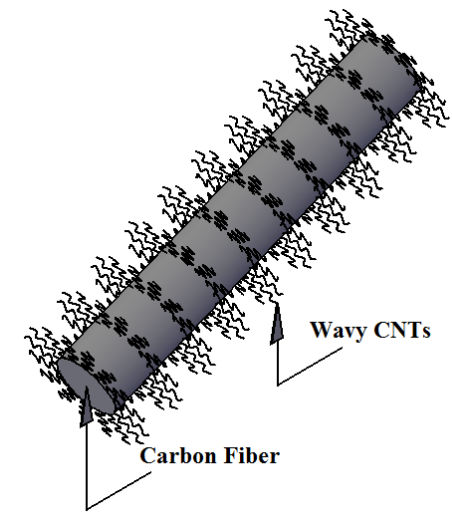

*Fig. 2 Fuzzy fiber with wavy CNTs radially grown on its circumferential surface*

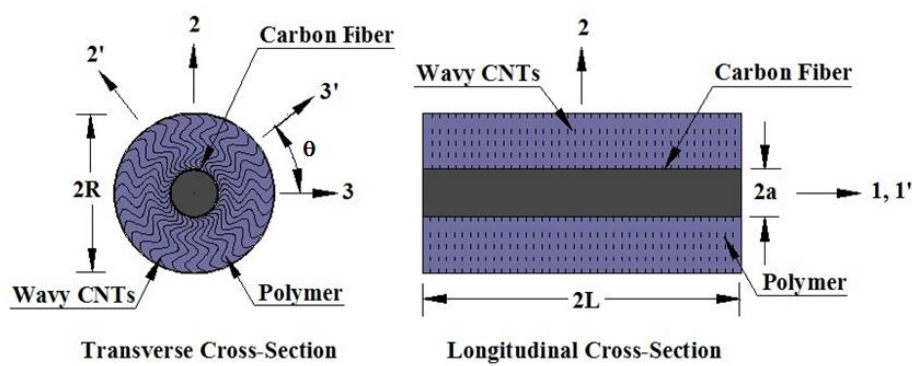

*(a) Cross sections of the CFF with wavy CNTs being coplanar with the 2–3 plane*

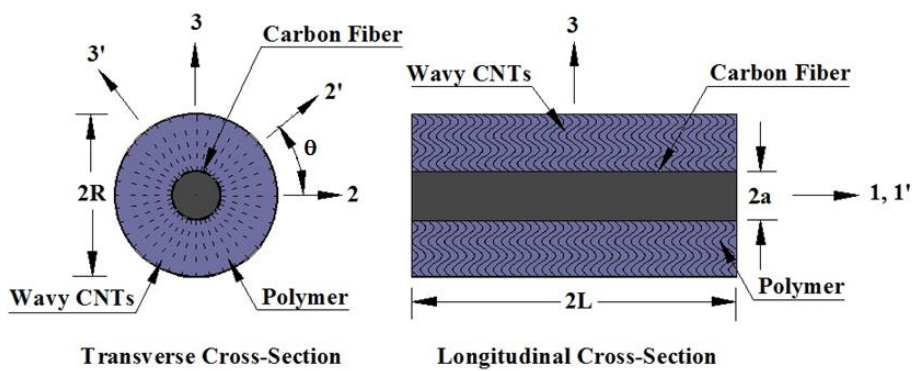

*(b) Cross sections of the CFF with wavy CNTs being coplanar with the 1–3 plane*

*Fig. 3 Transverse and longitudinal cross sections of the CFF in which wavy CNTs are coplanar with either the 2–3 (2'–3') or the 1–3 (1'–3') plane*

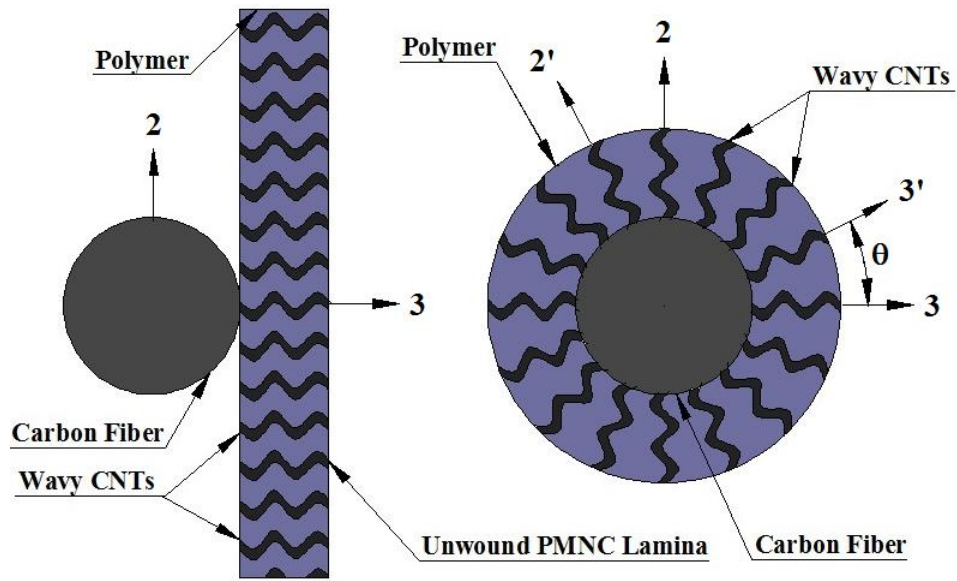

*Fig. 4 Transverse cross sections of the CFF with unwound and wound PMNC*

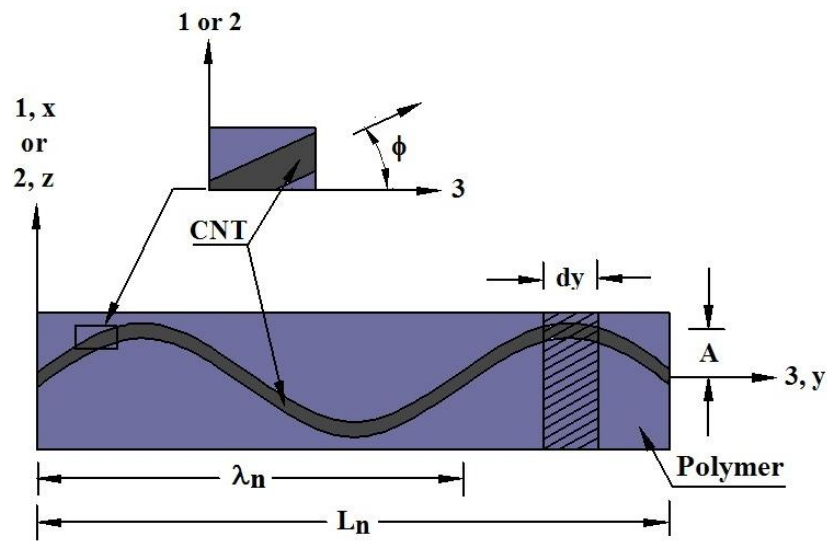

*Fig. 5 RVE of the unwound PMNC material containing a wavy CNT is coplanar with either the longitudinal plane (i.e., 1–3 or 1'–3' plane) or the transverse plane (i.e., 2–3 or 2'–3') of the carbon fiber*

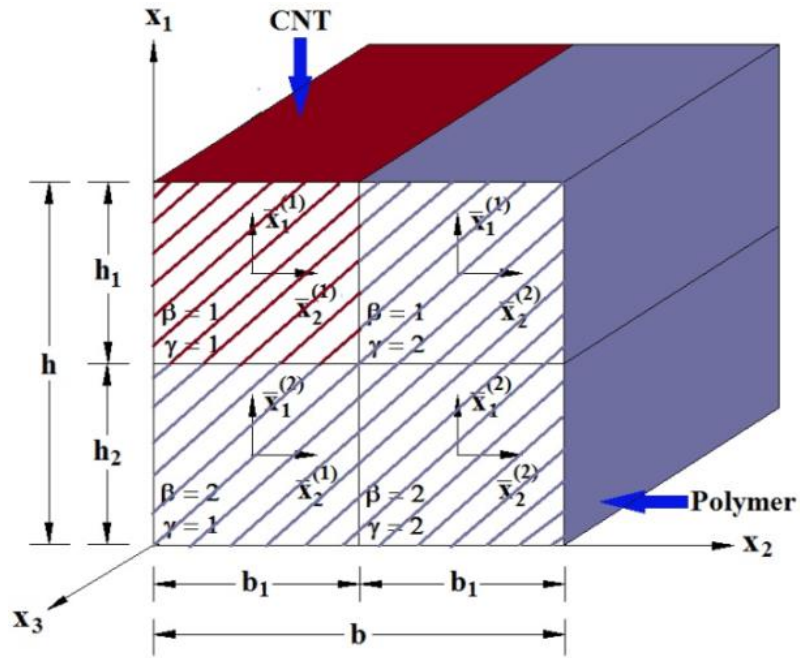

*Fig. 6 Repeating unit cell of the unwound PMNC material with four subcells ($\beta$, $\gamma$ = 1,2)*

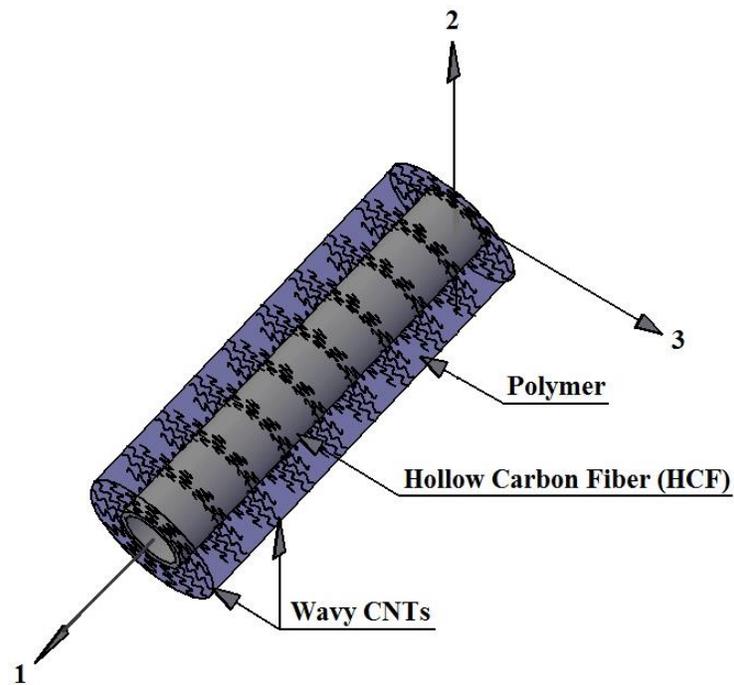

*Fig. 7 Schematic diagram of a novel FFHE*

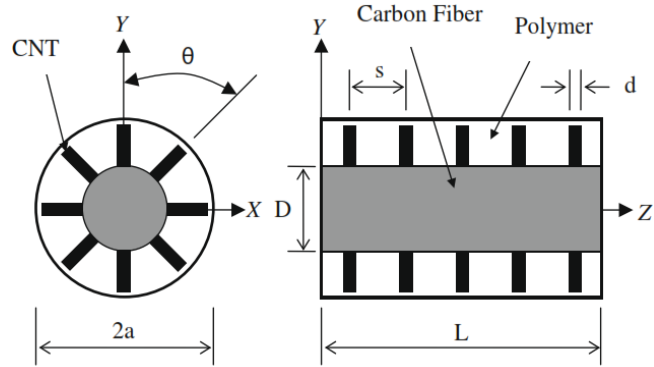

*Fig.. 8 Transverse and longitudinal cross sections of the NRLC (adapted from Ref. [11])*

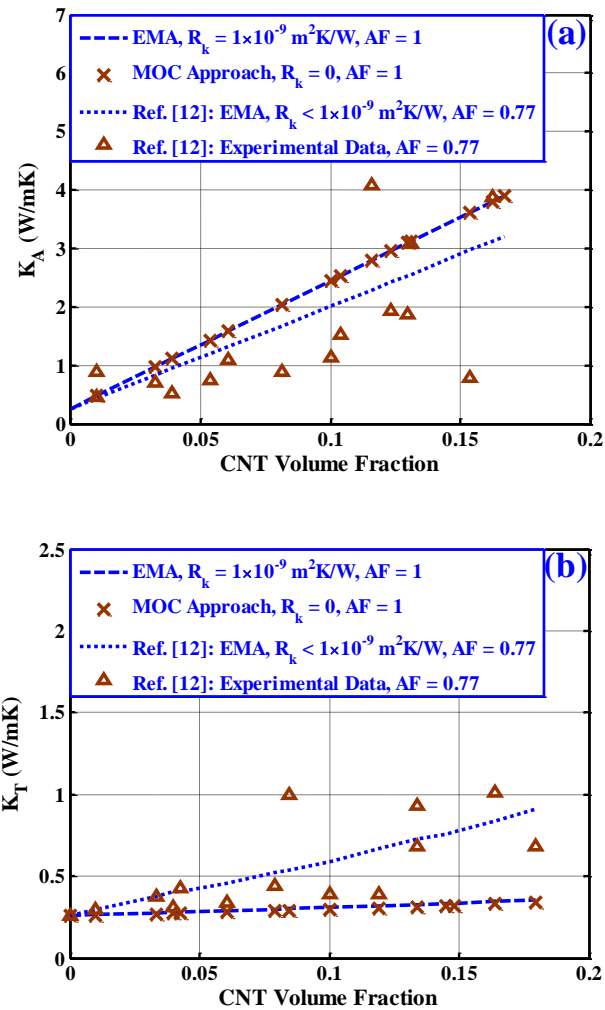

*Fig. 9 Comparisons of the effective (a) axial ($K_A$) and (b) transverse ($K_T$) thermal conductivities of the aligned CNT-polymer nanocomposite estimated by the MOC and the EM approaches with those of the experimental data [12]*

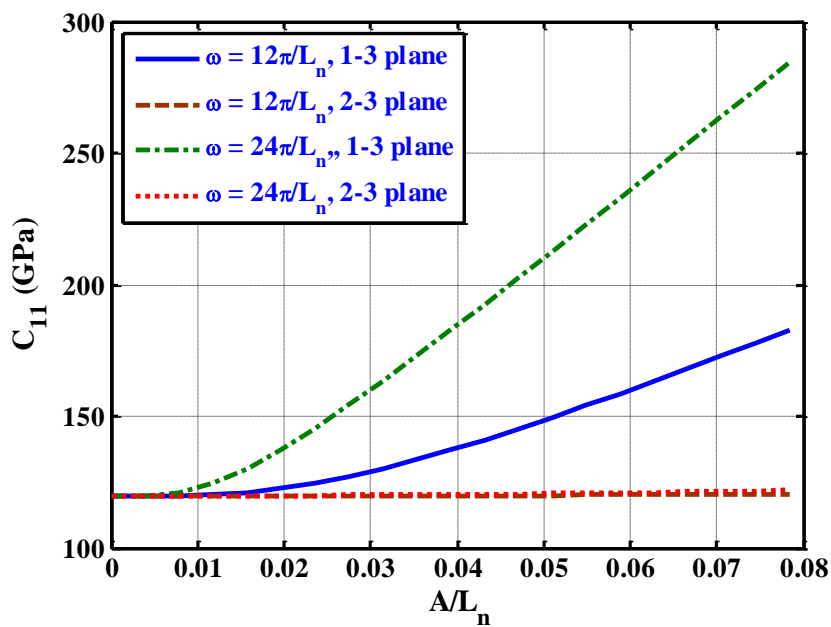

*Fig.10 Variation of the effective elastic coefficient $C_{11}$ of the FFRC with the waviness factor*

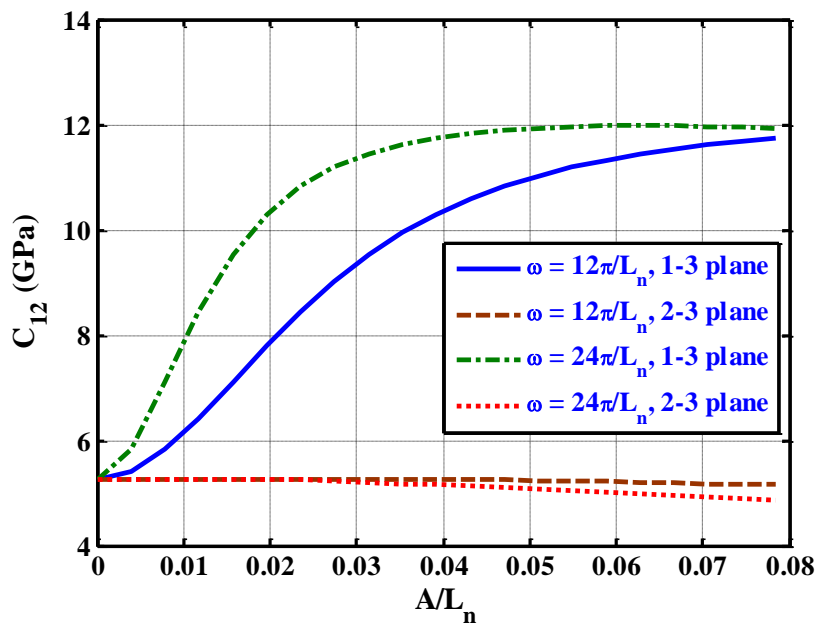

*Fig.11 Variation of the effective elastic coefficient $C_{12}$ of the FFRC with the waviness factor*

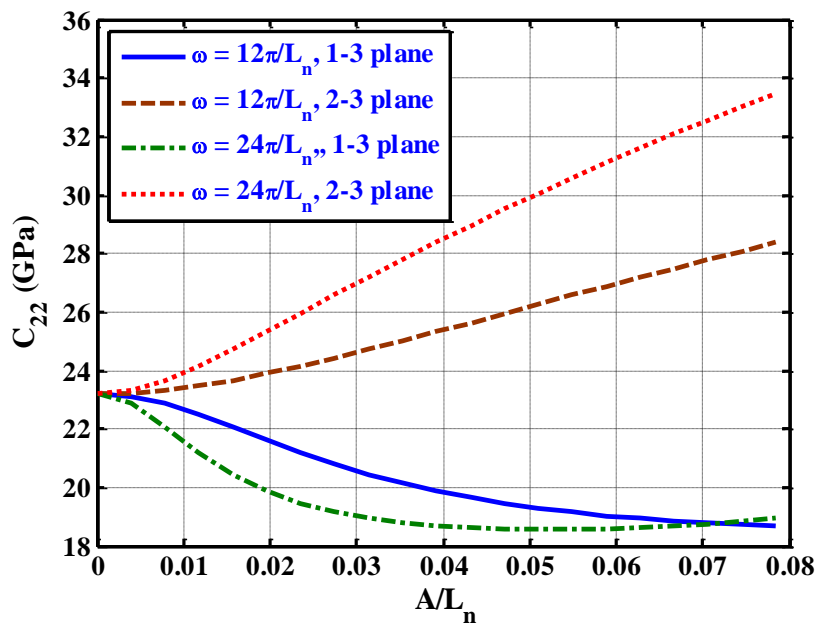

*Fig.12 Variation of the effective elastic coefficient $C_{22}$ of the FFRC with the waviness factor*

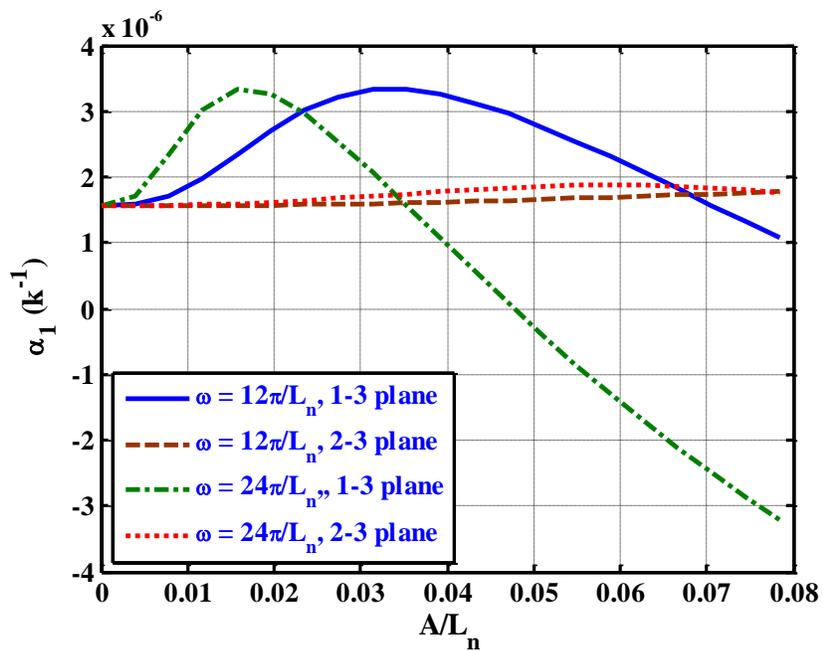

*Fig. 13 Variation of the axial CTE ($\alpha_1$) of the FFRC with the waviness factor ($\Delta T = 300K$)*

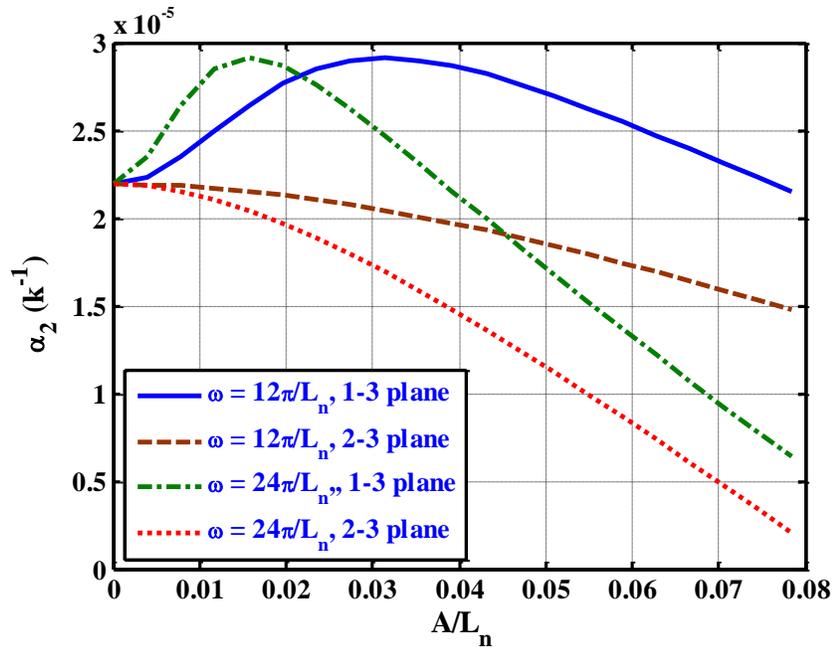

*Fig. 14 Variation of the transverse CTE ($\alpha_2$) of the FFRC with the waviness factor ($\Delta T = 300K$)*

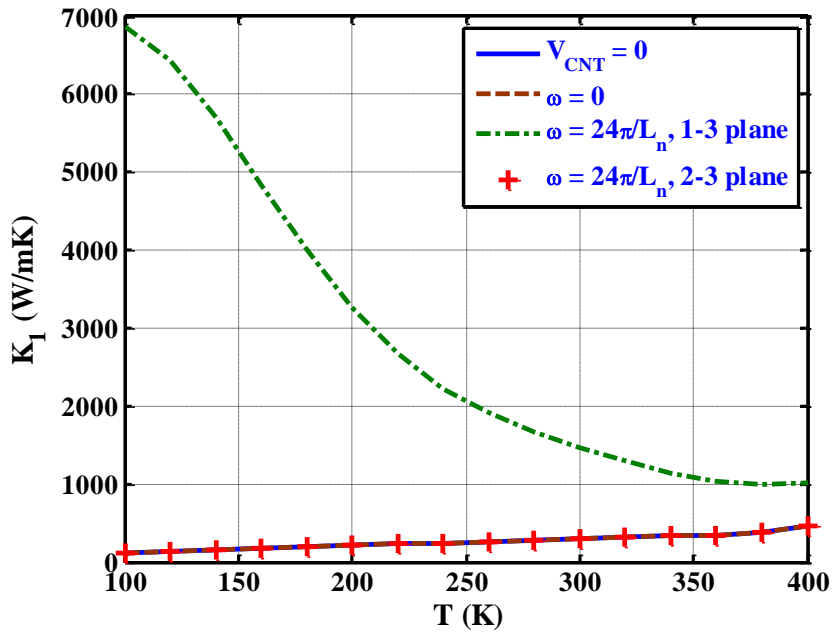

*Fig. 15 Variation of the effective axial thermal conductivity ($K_1$) of the FFRC with the temperature ($R_k = 0$)*

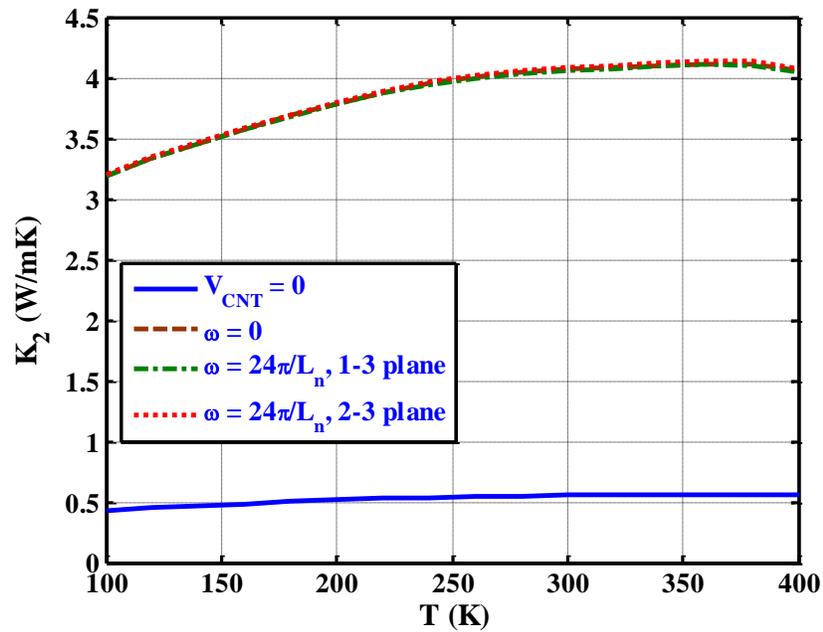

*Fig. 16 Variation of the effective transverse thermal conductivity ($K_2$) of the FFRC with the temperature ($R_k = 0$)*

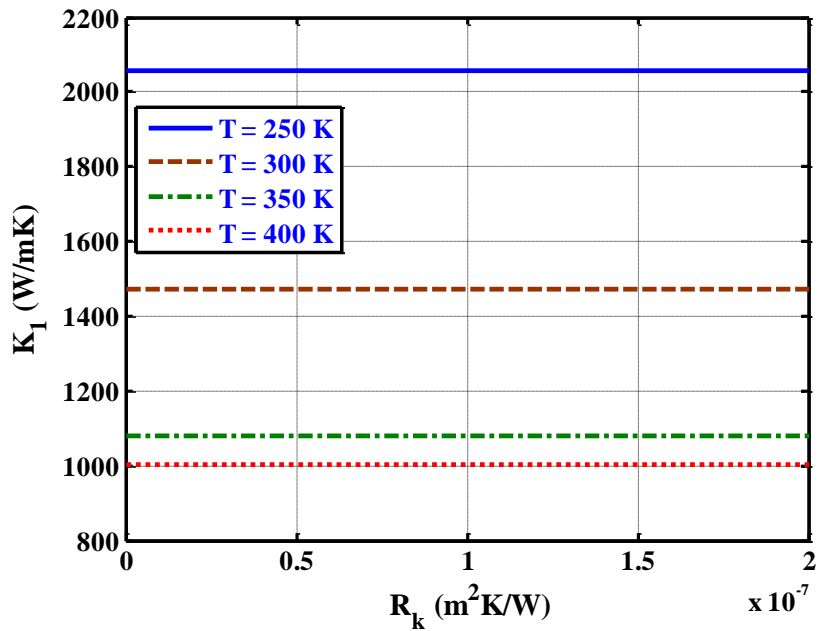

*Fig. 17 Variation of the effective axial thermal conductivity ($K_1$) of the FFRC with the CNT/polymer matrix interfacial resistance ($R_k$) when the wavy CNTs are coplanar with the 1–3 plane ($\omega = 24\pi/L_n$)*

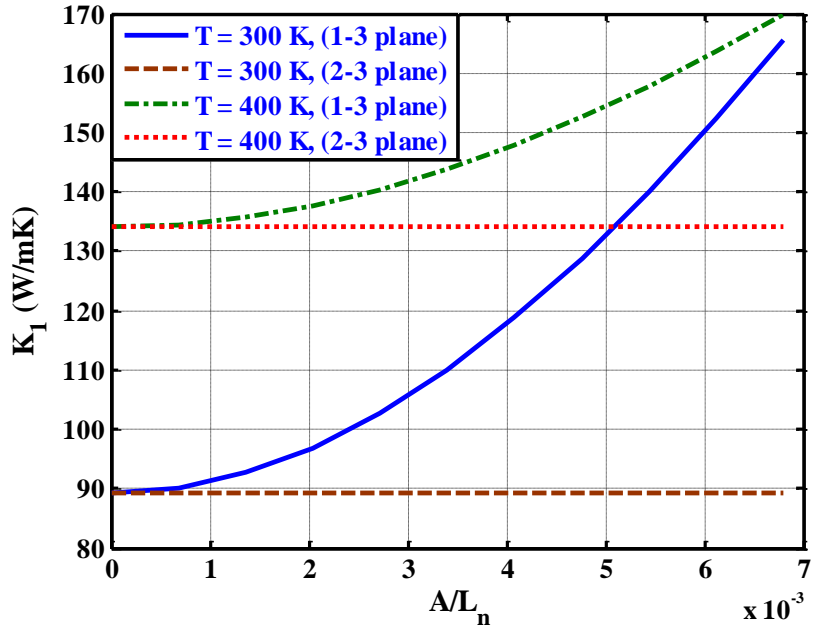

*Fig. 18 Variation of the effective axial thermal conductivity ($K_1$) of the FFHE with the waviness factor ($R_k = 0$, $\omega = 24\pi/L_n$)*

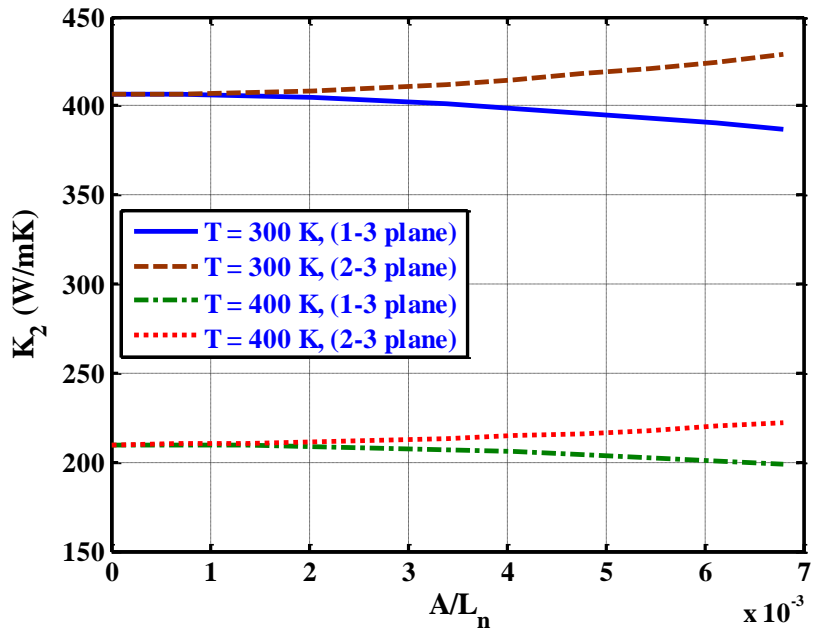

*Fig. 19 Variation of the effective transverse thermal conductivity ($K_2$) of the FFHE with the waviness factor ($R_k = 0$, $\omega = 24\pi/L_n$)*